\documentclass[12pt,preprint]{aastex}

\begin{document}

\title{Cepheid Masses:  FUSE Observations of S Mus
  \altaffilmark{1} }

\author{Nancy Remage Evans, }   
                                                                                     
\affil{Smithsonian Astrophysical Observatory, MS 4,  
 60 Garden St., Cambridge, MA 02138}                 

\author{Derck Massa}                                 
\affil{NASA's GSFC, SGT, Inc, Code 681.0, Greenbelt, MD 20771, USA} 
\author{Alexander Fullerton}                                        
\affil{Univ. of Victoria, Dept. of Physics and Astronomy, PO Box 3055, 
Station CSC, Victoria, BC, Canada}
\author{George Sonneborn and Rosina Iping}                             
\affil{NASA's GSFC,  Code 681, Greenbelt, MD 20771, USA}

%\centerline{email: nevans@cfa.harvard.edu}

\altaffiltext{1}{Based on observations made with the NASA-CNES-CSA 
 Far Ultraviolet Spectroscopic Explorer  Satellite.  FUSE is operated for 
 NASA by the Johns Hopkins University under NASA contract NAS5-32985}

%\altaffiltext{2}{York University, 4700 Keele St., North York, Ont. M3J 1P3 Canada} 

%\hfill\vfill
%\centerline{Address for Correspondence:}

%\centerline{Nancy R. Evans}

%\centerline{Smithsonian Astrophysical Observatory}

%\centerline{MS 4, 60 Garden St., Cambridge, MA 02138}

%\baselineskip=20pt

\begin{abstract}

   S Mus is the Cepheid with the hottest known companion. The
   large ultraviolet flux means that it is 
   the only Cepheid companion for which the
   velocity amplitude could be measured with the echelle mode of the HST GHRS.
   Unfortunately, the high temperature is difficult to constrain at
   wavelengths longer than 1200 \AA\/ because of the degeneracy between
   temperature and reddening. We have obtained a FUSE spectrum in order to
improve the    
determination of the temperature of the companion. Two
   regions which are temperature sensitive near 16,000 K but relatively
   unaffected by H$_2$ absorption (940 \AA, and the Ly $\beta$ wings)
   have been identified. By
   comparing FUSE spectra of
   S Mus B with spectra of standard stars, we have determined a temperature
   of 17,000 $\pm$ 500 K. 
   The resultant Cepheid mass is 6.0 $\pm$ 0.4 M$_\odot$.
This mass is consistent with main sequence evolutionary tracks 
with a moderate amount of convective overshoot.  

\end{abstract} 

\keywords{stars, clusters, X-rays, star formation}
%\keywords{globular clusters,peanut clusters,bosons,bozos}

\section{Introduction}

Observational  determinations of Cepheid masses are a 
long-standing goal both in order to have a thorough understanding 
of these primary distance indicators and also because they
provide an excellent benchmark for stellar evolutionary calculations.
The most important uncertainty in evolutionary tracks of massive
stars near the main sequence is the importance of core convective 
overshoot, which determines the lifetime on the main sequence and 
the luminosity in subsequent phases.  When the mass of a Cepheid 
can be measured, it can be combined with an accurate luminosity, 
and compared with theoretical predictions. 

Ultraviolet high resolution spectroscopy has provided a group of 
double-lined spectroscopic binaries containing a Cepheid.  Specifically,
the orbital velocity amplitudes of the hot companions of Cepheids
could be measured originally with IUE, and until recently with the Hubble 
Space Telescope (HST) Space Telescope Imaging Spectrograph (STIS) or 
Goddard High Resolution Spectrograph (GHRS).  This orbital velocity 
amplitude can be combined with the orbital velocity amplitude of the 
Cepheid from a ground-based orbit and the mass of the companion to 
produce the mass of the Cepheid.  Typically, a very accurate temperature 
or spectral type  for the hot companion can be  
obtained  from IUE low resolution spectra from  1200 to 3200 \AA,
from which a mass can be accurately inferred.  

For the S Mus system, the orbit of the Cepheid S Mus A has been 
determined several times with increasing accuracy as more data 
have been obtained (Evans, 1990; B\"ohm-Vitense et al. 1997; and
Petterson et al. 2004).

The hot companion of the Cepheid, S Mus B, is sufficiently bright 
at 1720~{\AA} that it could be observed with the echelle mode 
of HST/GHRS which provided a resolution of 80,000
(B\"ohm-Vitense et al. 1997). The orbital velocity amplitude of S Mus B they
found from two GHRS observations is 30.6 km s$^{-1}$ with an uncertainty of
5\%. The uncertainty is dominated by the centering of the star in the large
science aperture for the first observation. This is the most accurate velocity
amplitude measured for a  Cepheid companion.

The high temperature of S Mus B, however, means that 
its temperature (and hence its inferred mass) is 
less accurately determined than that for cooler companions in other 
systems.  For late B stars, the energy distribution 
turns over between 1200 and 1400 \AA, making that region of 
the spectrum extremely temperature sensitive.  For an early B 
star, the spectrum rises monotonically 
toward shorter wavelengths to the end of the IUE 
spectral range ($\sim$1200 \AA).  This means the effects of 
reddening and temperature are much more difficult to disentangle, 
and hence the temperature is less accurately determined.  
The reddening of the system is  E(B-V) = 0.21 mag (Evans, 
Massa and Teays 1994), which is large enough that it must 
be taken into account.

A number of approaches to determining the temperature in the 
wavelength range 1200 to 3200 \AA\/ have been used, as summarized
by B\"ohm-Vitense et al (1997), including energy 
distributions from IUE low resolution spectra, and
Si lines near 1300 \AA\/ from IUE high resolution spectra
(Evans, Massa and Teays, 1994).  In addition, two Voyager
spectra were obtained to extend the energy distribution to 
950 \AA\/ (Evans, Holberg and Polidan, 1996).  
The difficulty in interpreting these spectra comes from the 
heavy absorption by H$_2$ molecular absorption bands.  In the 
low resolution Voyager spectra, approximate corrections had 
to be incorporated to compensate for this absorption.   

As a substantial refinement to this basic approach, we obtained a                 
   high-resolution FUSE spectrum of S Mus B in order to determine its                
      temperature more precisely, with the ultimate goal of improving the               
         estimated mass of its Cepheid companion.

\section{{\it FUSE} Observations}

{\it FUSE} consists of four coaligned, prime-focus telescopes and 
Rowland-circle spectrographs that provide high-resolution ($\sim$15,000) 
spectra of the wavelength region between 905 and 1187~{\AA}.  To maximize 
throughput across this waveband, two of the telescope/spectrograph channels 
have SiC coatings to cover the range $\sim$905 -- 1105~{\AA}, while the 
other two have LiF coatings to cover $\sim$980 -- 1187~{\AA}. The spectra 
from all four channels are recorded simultaneously by two photon-counting 
detectors.  Details of the design and performance of these instruments are 
provided by Moos et al.~(2000) and  and Sahnow et al.~(2000), respectively.

The S~Mus system was observed by {\it FUSE} on 2002 March 2 as the only 
target in Guest Investigator program C011.  The observations were obtained 
through the {30\arcsec $\times$ 30\arcsec} (LWRS) aperture in time-tag (TTAG)
mode.  The total integration time was 16.26~ks, which was divided into four
exposures taken during four consecutive orbital viewing windows.  Thus, the 
observation spans the interval from 6:42 to 12:35 UT; or MJD 52335.2798 to 
52335.5242.

The photon lists from the four exposures were concatenated for each 
detector segment before the spectra were extracted and calibrated by 
means of CalFUSE version 2.2.3.  These processing steps included application 
of corrections for small, thermally-induced motions of the diffraction 
gratings; removal of extraneous counts due to ``event bursts"; removal of 
detector and scattered-light backgrounds; removal of thermal and electronic 
distortions in the detectors; correction of residual astigmatism in the 
spectrograph optics; extraction of a one-dimensional spectrum by summing 
over the astigmatic height of the two-dimensional image; correction for the 
minor effects of detector dead time; and application of flux and wavelength 
calibrations.  These manipulations produced four spectra from the two SiC 
channels, and four spectra from the two LiF channels, which are characterized
by signal-to-noise ratios of 20--25 in the stellar continuum.

In addition, archival {\it FUSE} spectra of the comparison stars listed in 
Table~1 were retrieved from MAST\footnote{The archiving of non-HST data at 
MAST is supported by the NASA Office of Space Science via grant NAG5-7584 
and by other grants and contracts.  STScI is operated by the Association of 
Universities for Research in Astronomy, Inc., under NASA contract NAS5-26555.}.
Since the comparison stars are bright, they were observed in histogram (HIST)
mode through the LWRS aperture, with substantially shorter integration times 
than was used for S~Mus.  All these data were processed uniformly with 
CalFUSE version 2.4.1 on an 
exposure-by-exposure basis, before being aligned and coadded for comparison 
with the far-UV spectrum of S~Mus.

\section{Comparisons}

The approach outlined in \S 1 is to use FUSE spectra of B-type stars with
well-determined temperatures to constrain the temperature of S Mus B. We
therefore sought spectral regions that are sensitive to temperature in the
spectral range of B3 V to B5 V, which are also lightly contaminated by H$_2$
absorption lines. For comparison purposes we used FUSE spectra of stars with
temperatures and gravities determined by one of us (DM) from Str\"omgren
photometry, using the calibration of Napiwotzki et al (1993), listed in Table
1. The observed spectra were $\it reddened$ to match the E(B-V) of S Mus
[E(B-V) = 0.23 mag].  We used the Fitpatrick (1999) $R(V) = 3.1$ curve,     
extrapolated into the Far-UV.  Such an extrapolation has been demonstrated   
by Sofia et al.\ (2005) to be a reasonable approximation for the Far-UV      
extinction.   They were
then overplotted on the S Mus spectrum using the flux in the wavelength range
1140 to 1170 \AA\/ to normalize them. Figs. 1 and 2 show two examples of the
comparisons for HD 51013 (B3 V) and HD 37332 (B5 V) respectively. The E(B-V)
of S Mus corresponds to a column density of $N_{H\,I}$ =
 1.4 x 10$^{21}$ cm$^{-2}$.
H$_2$ absorption for this column density has been overplotted to assess how
severely the stellar spectrum is affected in different wavelength regions.
Standard values for the ISM provided a column density of H$_2$ a little less
than 10$^{20}$cm$^{-2}$, for example Evans et al. (1996).  

In this temperature range two regions were identified from these 
comparisons as being both temperature sensitive and relatively
unaffected by H$_2$: (1) the wings of Ly$\beta$ ($\simeq$ 1010 to
1040 \AA) and (2) the two shortest wavelength regions with measurable
flux, 945 and 960 \AA.  The core of Ly$\beta$ is strongly absorbed
by H$_2$, but the wings are only lightly absorbed. The wings of  Ly$\beta$
 are gravity sensitive, but we have restricted our comparisons
to main sequence stars, so this should not be a problem.

Enlarged versions of these regions of the same two standard stars (HD 51013
and HD 37332) are presented in Figs. 3 and 4 with the spectrum of S Mus and an
H$_2$ absorption template overplotted.

All 4 figures show that S Mus B is a very good match to 
HD 51013 (B3 V) and distinctly different from HD 37332 (B5 V) in
the temperature sensitive wavelength regions in Figs. 3 and 4.  
Figures such as 1 and 2 were examined for all the stars in Table 1.
S Mus B is also clearly different (hotter) than HD 35899 (B5 V) 
in these regions.
Based on the temperatures for the two stars in Table 1 which 
most closely match S Mus B (HD51013, T = 17,100K, and 
HD 35899, T = 16,700K), we adopt T = 17,000K for S Mus B 
with an uncertainty of $\pm$ 500K. 

Subsequent to the FUSE S Mus observation 
we have been obtaining FUSE spectra of the massive stars in 
eclipsing binaries (Andersen, 1991) in order to determine their
temperatures by the same approach (Evans et al., 2005).  We have 
come to realize that evolution beyond the zero age main sequence
results in enhanced contrast in lines.  We suspect there 
may be a modest amount of evolution in S Mus B based on the line 
strength in Figs 1-4.  We will revisit
the S Mus spectrum when we have completed the analysis of the 
eclipsing binaries.    

\section{Discussion}

In order to determine the mass of S Mus B corresponding to this           
   temperature, we used masses from the compilation
by Andersen (1991) derived from very accurate eclipsing binary 
solutions.  We have combined these with  recent temperatures
from Ribas et al. (2000). These temperatures are based on 
Str\"omgren photometry, and should be comparable to the temperatures
of the standard stars. In Fig. 5 we show the relation between 
temperature and mass for the stars in the Anderson list more 
massive than 2.5 M$_\odot$ (O and B stars).  In order to obtain a mass 
corresponding to the temperature of S Mus B, we did a linear fit to 
the data for $\log T_{\rm eff}$  4.27 to 4.17  (T 18,500 to 14,800 K).  The resulting
mass for 17,000 K is 5.3 M$_\odot$.  A change of 500 K results in a 
change of 0.26 M$_\odot$.

The orbital velocity amplitude ratio of the Cepheid to the hot companion 
was found to be 1.14 $\pm$ 0.06 from the GHRS echelle observations 
of the companion and the ground-based Cepheid orbit (B\"ohm-Vitense
et al. 1997).  Combining this with the companion mass found here
results in a Cepheid mass of 6.0 $\pm$ 0.4 M$_\odot$.  This value 
only differs slightly from the previous determination
(5.9 $\pm$ 0.7 M$_\odot$ B\"ohm-Vitense et al., 1997)
but the error bars are significantly reduced.  

This temperature and mass determination supersedes previous estimates, because
it lifts the degeneracy in the energy distribution longward of 1200 {\AA} and
avoids the need for the approximate corrections for H$_2$ absorption required
to interpret Voyager spectra.

Fig. 6 summarizes the information currently available for Cepheid 
masses.  (Luminosities are taken from Evans et al, 1998).
For comparison, the luminosity predicted for the tips of the
blue loops the evolutionary tracks 
 from several several groups is shown.  The two lines in 
the center are from the Padua and Geneva groups for moderate 
overshoot.  To the left is prediction from the Padua tracks for 
their maximum overshoot.  The line on the right is from Becker (1981)
with no overshoot.  The mass we have determined for S Mus 
clearly favors moderate overshoot.

\section{Summary}
We have used a FUSE spectrum of the hot companion of the Cepheid
S Mus to determine the temperature of S Mus B. By combining the mass 
corresponding to the temperature with the previously determined 
orbital velocity ratio of the two stars results in a mass of the 
Cepheid of 6.0 $\pm$ 0.4 M$_\odot$.

Acknowledgments
It is a pleasure to thank FUSE staff for assistance in obtaining 
this data.  Comments from an anonymous referee improved the discussion.
This research was supported by NASA FUSE grant  NAG5-11946
(to NRE) and  Chandra X-ray  Center NASA Contract NAS8-03060

\clearpage  

%\begin{table}
\begin{deluxetable}{lrlrrlccc}
\footnotesize
\tabletypesize{\scriptsize}
\tablecaption{ Comparison Stars
\label{tbl-1}}
\tablewidth{0pt}
\tablewidth{6truein}
%\caption{Table 1: Comparison Stars}
%\tableline
%\begin{tabular}{llccccccc}
\tablehead{
\colhead{ Star} & \colhead{FUSE} &  \colhead{PI} & \colhead{Date} & \colhead{Exptime}  
& \colhead{ Spectral} & \colhead{ E(B-V)} &
\colhead{ T} & \colhead{log g} \\
\colhead{ } & \colhead{Data Set} & \colhead{ } & \colhead{ } & \colhead{ (sec)}
& \colhead{ Type} & \colhead{(mag) } &
\colhead{ (K)} & \colhead{cm s$^{-2}$} \\
}
% & & & & \\
% Star & Spectral & E(B-V) & T & log g \\
% & Type & (mag) & (K) & cm s$^{-2}$ \\
% & & & & \\
%\tableline
\startdata
HD 97991 & P1013001 &  Savage &  2000-05-29  &  59 & B1 V & 0.02 & 25,500 & 4.00 \\
HD 121800 & P1014401 & Savage &  2000-03-17  &  3987 & B1.5 V & 0.07 & 20,300 & 1.30 \\
HDE 233622 & P1012102 & Savage & 2000-01-12  & 4662 & B2 V & 0.03 & 21,000 & 3.60 \\
HD 51013 & A0630902 & Witt & 2001-04-05 & 4080 & B3 V & 0.00 & 17,100 & 4.30 \\
HD 261878 & P1310301 &  Friedman &  2001-02-20 &  1572 & B3 V & 0.04 & 15,500 & 4.50 \\
HD 72350 & A1290305 &  Nichols &  2001-05-15 & 806 & B4 IV & 0.16 & 15,650 & 3.00 \\
HD 35899 & B0600101 &  Fitzpatrick &  2001-01-03  &  972 & B5 V & 0.03 & 16,700 & 4.30 \\
HD 37332 & B0601201 &  Fitzpatrick &  2001-01-05 & 2309 & B5 V & 0.02 & 15,600 & 4.50 \\
HD 37641 & B0601501 &  Fitzpatrick &  2001-01-03 & 4011 & B6 V & 0.05 & 12,800 & 4.40 \\
\enddata

\end{deluxetable}
% \end{tabular}
% \end{table}

\clearpage

\figcaption[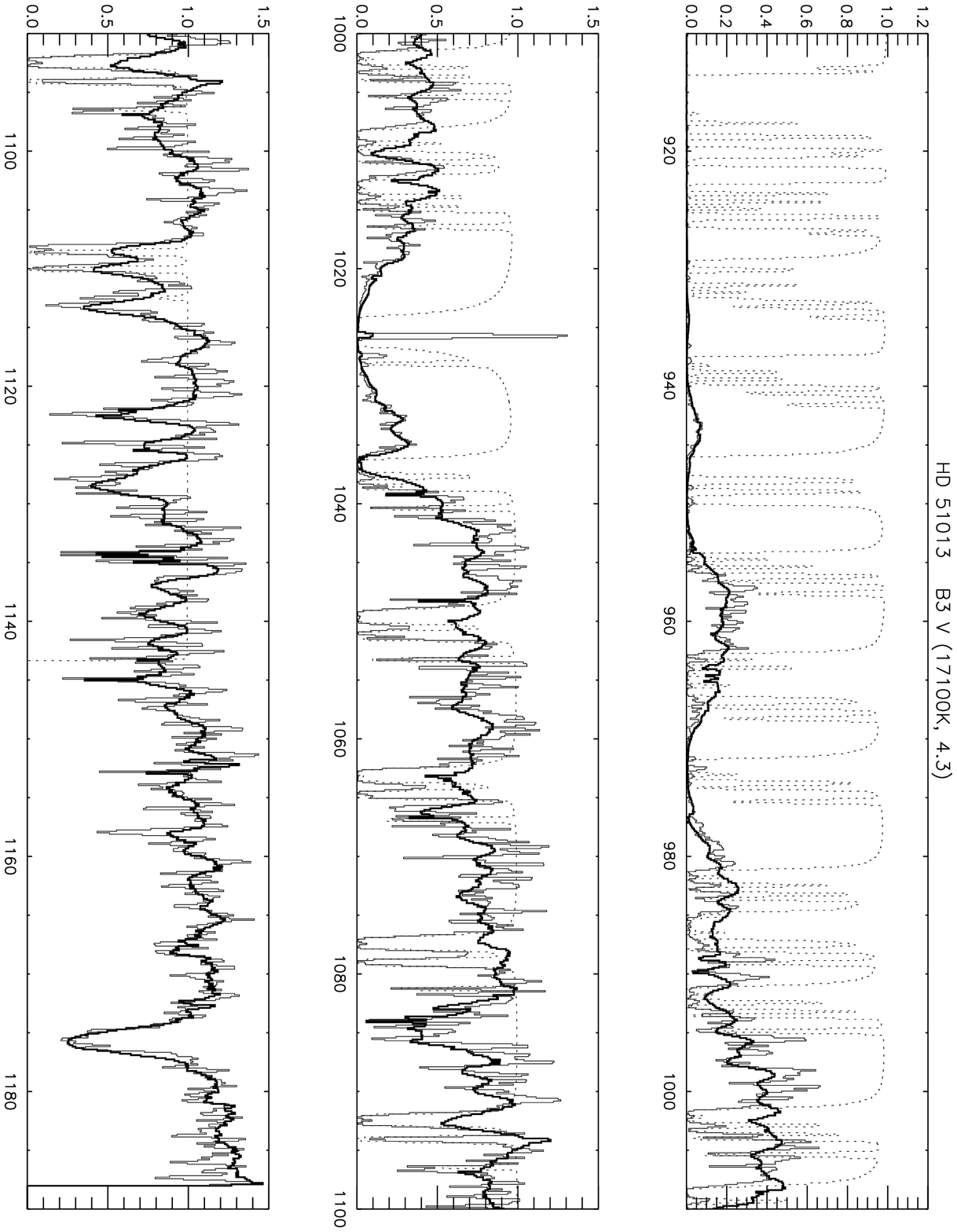]{S Mus compared with HD 51013 (B3 V).       
The lighter, noisier line is the S Mus spectrum.  The dotted line 
is the H$_2$ absorption spectrum.  The wavelength is in $\AA$; 
the flux has been normalized.
   \label{fig1}}

\figcaption[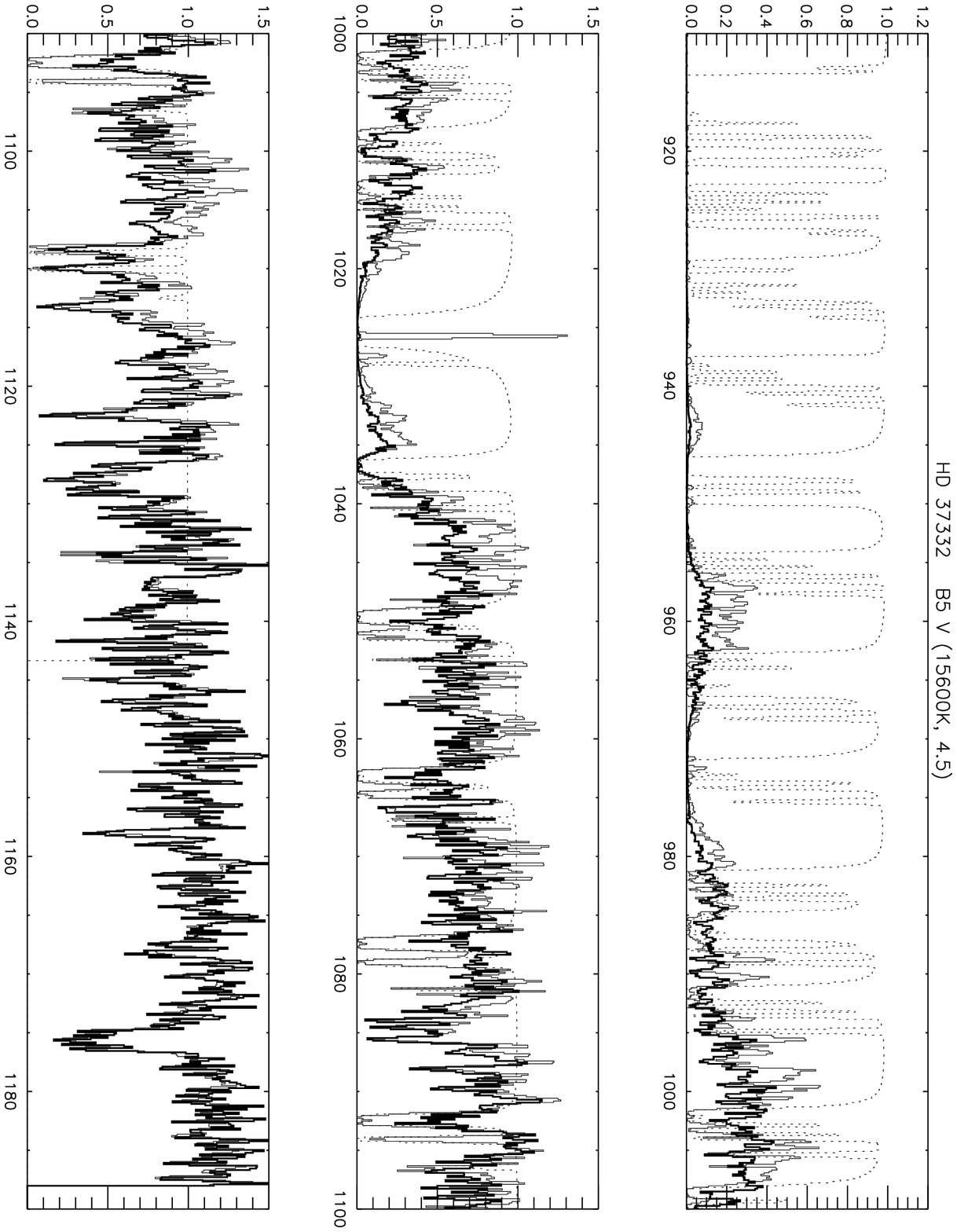]{S Mus compared with HD 37332 (B5 V).       
Line types are the same as in Fig. 1.  The wavelength is in $\AA$;     
the flux has been normalized. 
   \label{fig2}}

\figcaption[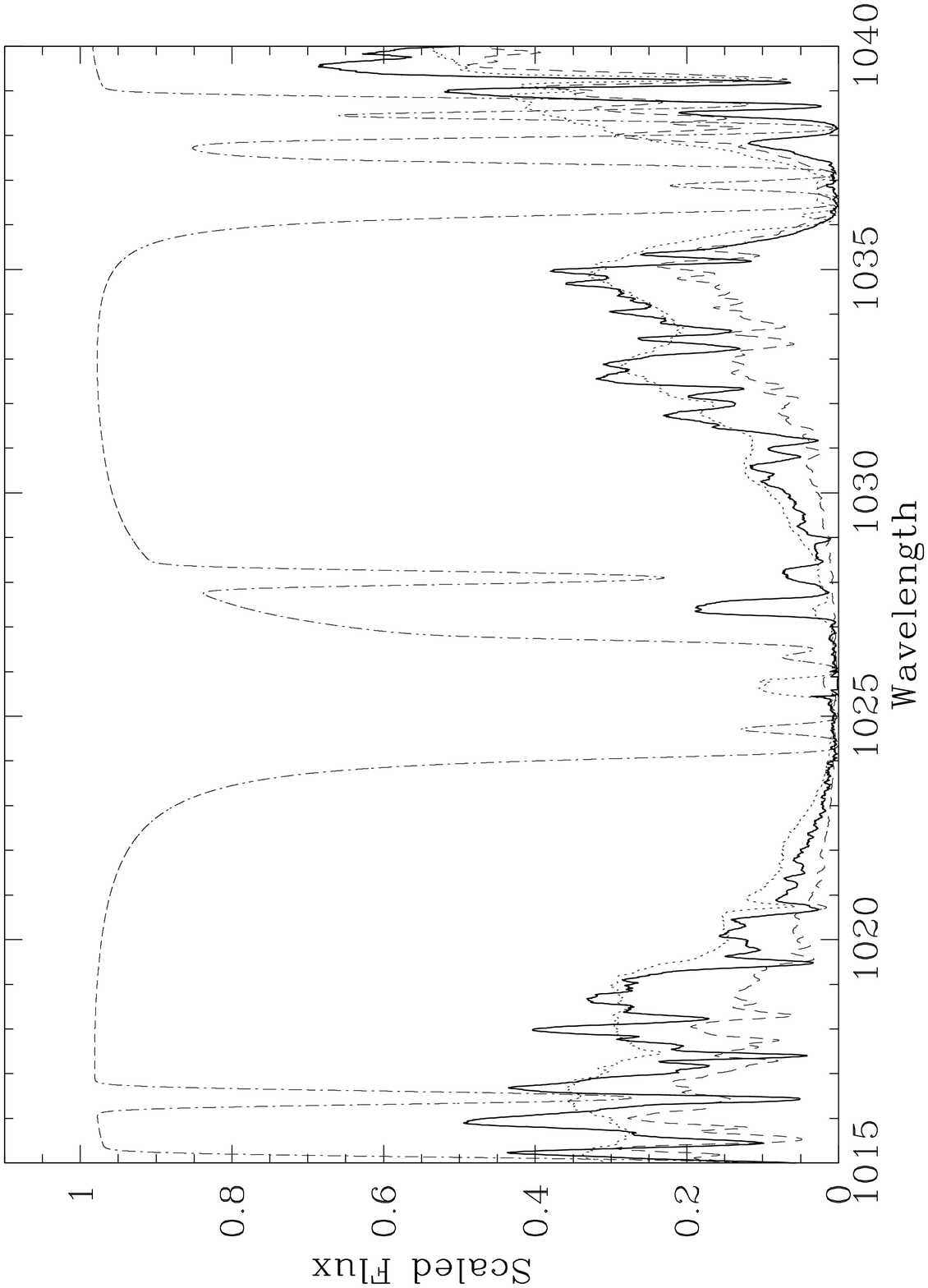]{The region near Ly$\beta$.  S Mus is the solid line;
the B3 V star is the dotted line; the B5 V star is the dashed line;
the H$_2$ absorption is the dot-dash line. A narrow airglow line in the center
of Ly$\beta$ has been excised.  The wavelength is in $\AA$.
   \label{fig3}}
   
\figcaption[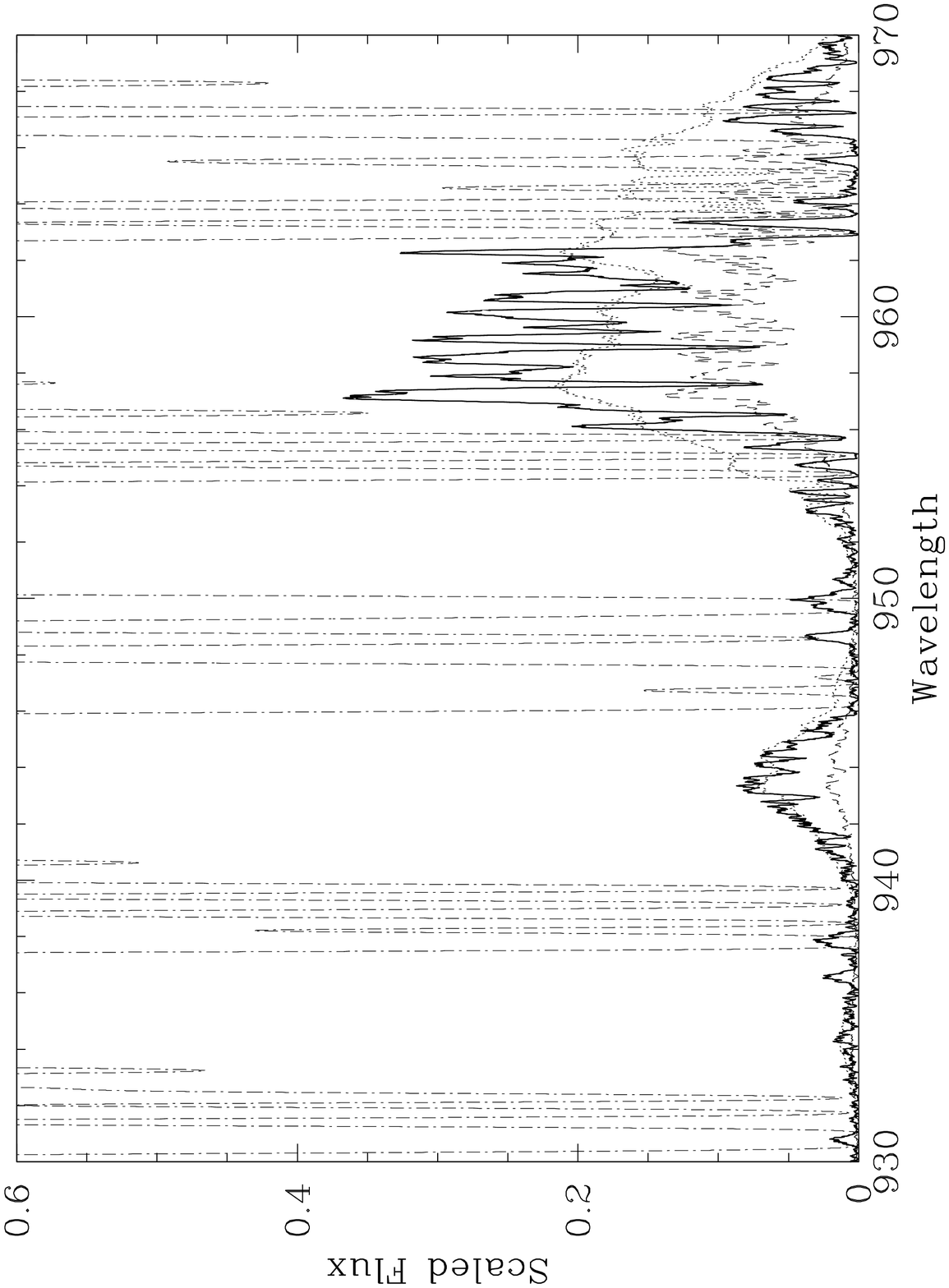]{Spectral comparisons in the 950 \AA\/ region.
The symbols are the same as in Fig. 3. The wavelength is in $\AA$.
   \label{fig4}}

\figcaption[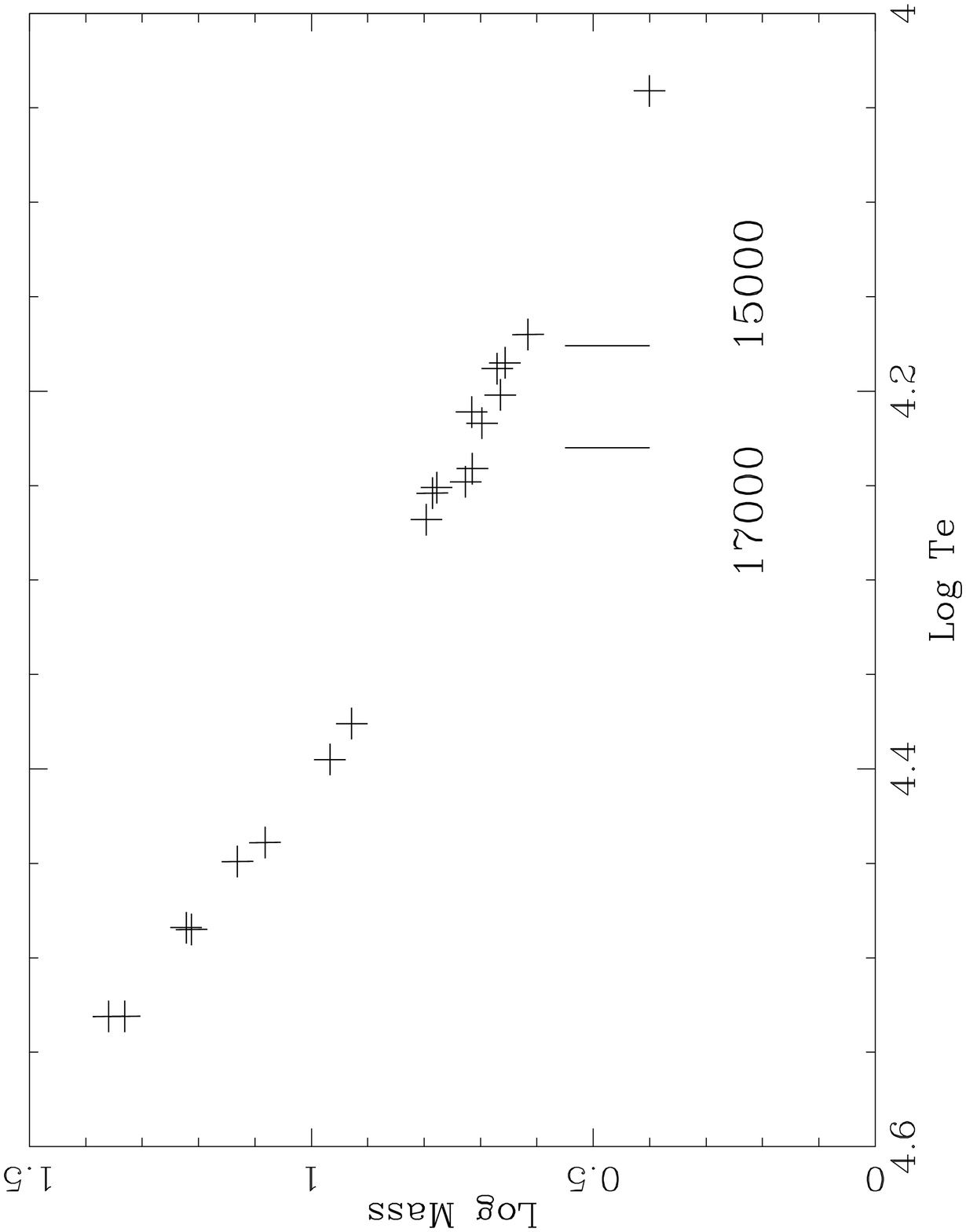]{The mass-temperature relation for  O and B
stars from the list of eclipsing binaries of Andersen and temperatures
from Ribas, et al. Mass is in M$_\odot$; temperature is in K.
   \label{fig5}}

\figcaption[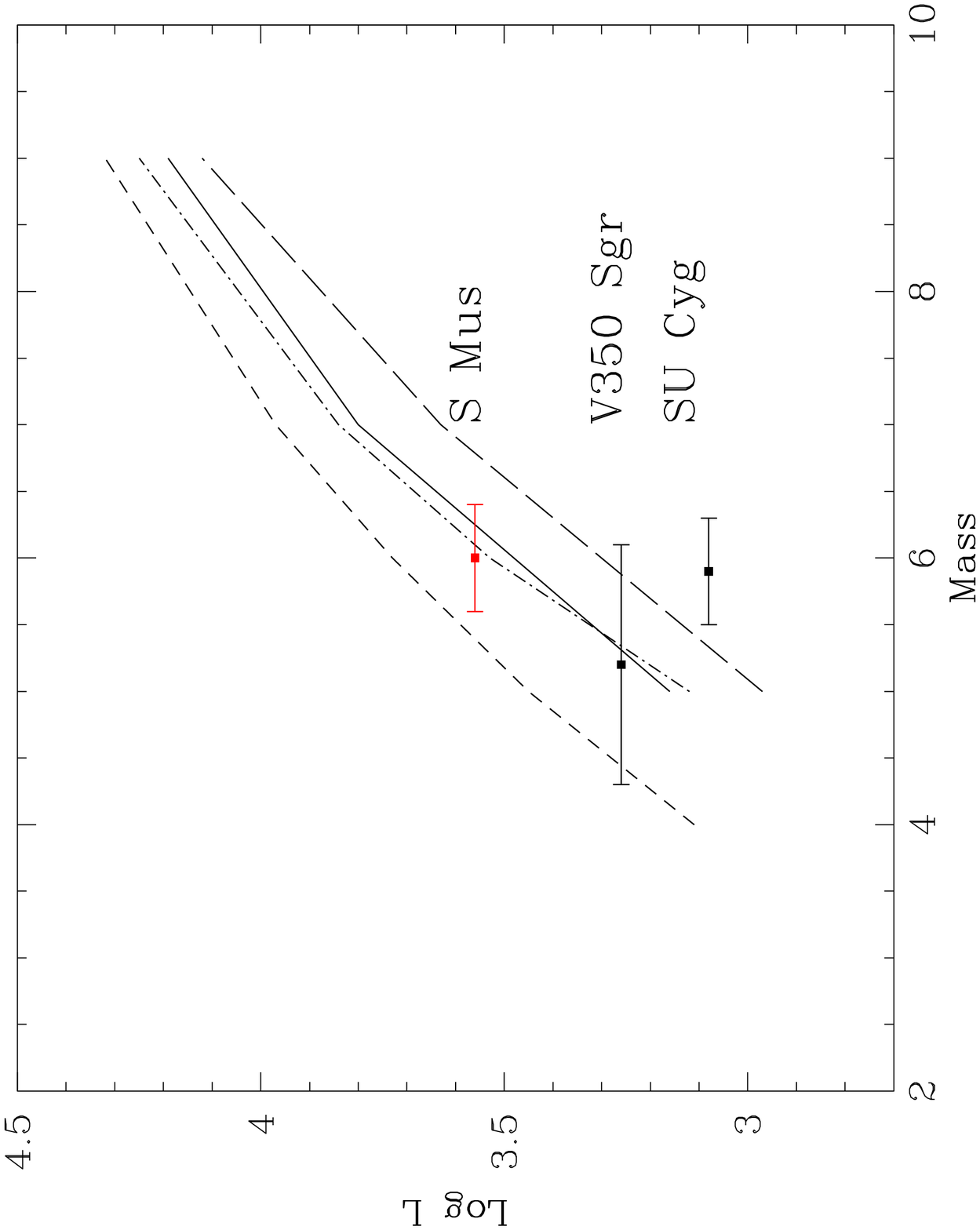]{Measured Cepheid  masses. For comparison, 
theoretical predictions from evolutionary tracks are shown with 
decreasing convective overshoot from left to right. Dashed line:
Bertelli, et al. 1986; dot-dash line: Bertelli, et al. 1994;
solid line: Schaller, et al. 1992; dashed line: Becker, 1981)
Mass and luminosity are in solar units.
   \label{fig6}}

%\plotone{hd51013_2.ps}
\plotone{f1.ps}

%\plotone{hd37332_1.ps}
\plotone{f2.ps}
%\plotone{lyb.ps}
\plotone{f3.ps}
%\plotone{950.ps}
\plotone{f4.ps}
%\plotone{ecbintmp.ps}
\plotone{f5.ps}
%\plotone{smusma.ps}
\plotone{f6.ps}
%\plottwo{spececga.ps}{spececgb.ps}

\end{document}